\documentclass[aps,twocolumn,preprintnumbers,superscriptaddress,floatfix,prl,10pt]{revtex4-2}
\usepackage{bbm}
\usepackage{bm}
\usepackage{amsmath}
\usepackage{amssymb}
\usepackage{empheq}
\usepackage{graphicx}
\usepackage{mathrsfs}
\usepackage{amsfonts}
\usepackage{amsthm}
\usepackage{color}
\usepackage{multirow}
\usepackage{bigints}
\usepackage{txfonts}
\usepackage{float}
\usepackage{hyperref}
\hypersetup{
     unicode=false,              
     pdftoolbar=true,           
     pdfmenubar=true,         
     pdffitwindow=false,      
     pdfstartview={FitH},    
     pdftitle={My title},       
     pdfauthor={Author},     
     pdfsubject={Subject},   
     pdfcreator={Creator},   
     pdfproducer={Producer}, 
     pdfkeywords={keyword1} {key2} {key3}, 
     pdfnewwindow=true, 
     colorlinks=false,       
     linkcolor=red,           
     citecolor=green,        
     filecolor=magenta,    
     urlcolor=cyan           
}

\makeatletter \tolerance = 10000 \tolerance = 10000

\makeatother
\begin{document}

\title{Hermitian chiral boundary states in non-Hermitian topological insulators}

\author{C. Wang}
\email[Corresponding author: ]{physcwang@tju.edu.cn}
\affiliation{Center for Joint Quantum Studies and Department of Physics, 
School of Science, Tianjin University, Tianjin 300350, China}
\author{X. R. Wang}
\email[Corresponding author: ]{phxwan@ust.hk}
\affiliation{Physics Department, The Hong Kong University of Science 
and Technology (HKUST), Clear Water Bay, Kowloon, Hong Kong}
\affiliation{HKUST Shenzhen Research Institute, Shenzhen 518057, China}

\date{\today}

\begin{abstract}
Eigenenergies of a non-Hermitian system without parity-time symmetry are complex in general. Here, we show that the chiral boundary states of higher-dimensional (two-dimensional and three-dimensional) non-Hermitian topological insulators without parity-time symmetry can be Hermitian with real eigenenergies under certain conditions. Our approach allows one to construct Hermitian chiral edge and hinge states from non-Hermitian two-dimensional Chern insulators and three-dimensional second-order topological insulators, respectively. Such Hermitian chiral boundary channels have perfect transmission coefficients (quantized values) and are robust against disorders. Furthermore, a non-Hermitian topological insulator can undergo the topological Anderson insulator transition from a topological trivial non-Hermitian metal or insulator to a topological Anderson insulator with quantized transmission coefficients at finite disorders.
\end{abstract}

\maketitle

\emph{Introduction.$-$}Topological states~\cite{thouless_prl_1982,haldane_prl_1988,wen_advphy_1995,hasan_rmp_2010,qi_rmp_2011,bansil_rmp_2016,beenakker_rmp_2015,chiu_rmp_2016,wen_rmp_2017,wabenalcazar_science_2017} are new types of states of matter that have attracted great attention from people working in various fields of physics, including electronic structures~\cite{kane_prl_2005,bernevig_science_2006,konig_science_2007,change_science_2013}, mechanics~\cite{kane_natphys_2014}, magnetics~\cite{xswang_prb_2017,ysu_prb_2017}, photonics~\cite{fmdhaldane_prl_2008,mhafezi_natphys_2011}, quantum information~\cite{kitaev_annphys_2003}, and ultracold atomic gases~\cite{ibloch_natphys_2012}. These states are featured by the robust chiral or helical boundary (surface, edge, or corner) states~\cite{hasan_rmp_2010,qi_rmp_2011,bansil_rmp_2016,beenakker_rmp_2015,chiu_rmp_2016,wen_rmp_2017,wabenalcazar_science_2017}. Unlike non-topological boundary states that are fragile, topological boundary states are guaranteed by the bulk-boundary correspondence rooted in the Stokes-Cartan theorem. Characterized by non-zero topological numbers, bulk bands can be gapped in topological insulators and topological superconductors or gapless in Weyl semimetals and Dirac semimetals. For example, Chern insulators (CIs), also known as the topological magnetic insulators, have bulk energy gaps and in-gap one-dimensional chiral edge channels with quantized conductances that are immune from disorders~\cite{eprodan_prl_2010,sliu_prl_2016,cliu_natmat_2020}.  
\par

Very recently, considerable research activities have focused on exploring new physics in non-Hermitian systems. Many theoretical and experimental works show that non-Hermicity gives rise to different phenomena that do not occur in a Hermitian system~\cite{nelson_prl_1996,makris_prl_2008,lee_prl_2016,xu_prl_2017,kunst_prl_2018,gong_prx_2018,yokomizo_prl_2019,kawabata_prx_2019,chlee_prl_2019,liu_prb_2021,yao_prl_2018,
zhou_prb_2018,luo_prl_2021,bbahari_science_2017,gharari_science_2018,bandres_science_2018,kkawabata_nc_2019,rhamazaki_prl_2019,cwang_prb_2020}. For instance, Poisson distribution, universal level-spacing statistics of localized states in Hermitian systems, becomes quasi-particle level spacing statistics in metallic phases of non-Hermitian systems~\cite{rhamazaki_prl_2019,cwang_prb_2020}. Moreover, a modified bulk-boundary correspondence is established to describe the non-Hermitian topological states of Su-Schrieffer-Heeger models~\cite{lee_prl_2016,kunst_prl_2018,yao_prl_2018,zhou_prb_2018}, CIs~\cite{bbahari_science_2017,gharari_science_2018,bandres_science_2018}, and quantum spin Hall insulators~\cite{kkawabata_nc_2019}. So far, non-Hermitian systems with parity-time (PT) symmetries are known to have real-energy spectra in PT-unbroken phases~\cite{cmbender_prl_1998,hyang_prl_2018,kkawabata_prr_2020}. However, whether other non-Hermitian systems can also have certain states whose effective Hamiltonian is {\it Hermitian} remains unclear.
\par 

The answer to this fundamental issue is partially resolved in one dimension (1D). People showed the existence of Hermitian localized topological states (end states) in 1D systems~\cite{lee_prl_2016,kunst_prl_2018,yao_prl_2018,zhou_prb_2018}. However, whether the results in 1D systems can be generalized to higher-dimensional systems is not clear at all and appears to be quite non-trivial. The Hermitian boundary states in 1D and higher dimensions are fundamentally different. Boundary topological states in 1D are localized and cannot transport particles, energy, or information, while topological edge and hinge states in two-dimensional (2D) and three-dimensional (3D) systems are extended and capable of transporting energy and information. Here, we present a systematic approach for achieving Hermitian chiral boundary states in bulk non-Hermitian topological phase. An effective Hermitian Hamiltonian describes such chiral boundary states while the bulk Hamiltonian is non-Hermitian. This can occur to chiral edge states in 2D CIs, chiral hinge states in 3D second-order topological insulators (3DSOTIs), and topological Anderson insulators (TAIs). Same as those in Hermitian systems, the transmission coefficient of each chiral channel in non-Hermitian systems is precisely 1, in contrast to the non-quantized value of a general non-Hermitian topological insulator~\cite{philip_prb_2018,chen_prb_2018,groenendijk_prr_2021,yi_arXiv_2021}. Besides, we verify the robustness of these Hermitian boundary states against impurities. 
\par

\emph{A generic approach.$-$}We first present a generic picture about how to realize 
Hermitian chiral boundary states out of a non-Hermitian topological insulator. 
The effective Hamiltonian of chiral boundary states of momentum $\pm p_1$ on opposite 
sides of a 2D topological insulator can be obtained by projecting total Hamiltonian $h$ 
($h\neq h^\dagger$) into the subspace spanned by two specific chiral boundary states 
$\psi_{\pm}(p_1)$ localized at two opposite edges~\cite{mkonig_jps_2008,bzhou_prl_2008,
jlinder_prb_2009}. For large enough sizes, the low-energy effective Hamiltonian 
$h^{\text{s}}(p_1)$ of chiral boundary states of well-defined momentum 
$p_1$ can be written as (to the linear order of $p_1$)
\begin{equation}
\begin{gathered}
h^{\text{s}}(p_1)=(t p_1+i\kappa)\sigma_3+i\kappa_0 \sigma_0.
\end{gathered}\label{eq1}
\end{equation}
Here, $\sigma_i$ with $i=0,1,2,3$ are the two-by-two identity matrix 
and three Pauli matrices acting on the subspace spanned by chiral 
boundary states $\{|\psi_{+}(0)\rangle,|\psi_{-}(0)\rangle\}$. 
$t=\hbar v_F$ with $v_F$ being the Dirac velocity of chiral boundary 
states. 
\par

In Eq.~\eqref{eq1}, the Hermitian term $tp_1\sigma_3$ describes 
two chiral edge channels that are localized at opposite edges of a 
sample and propagate along the $x$ direction. For samples of size 
$L$ much much larger than the width $\xi$ of edge states, there 
are no off-diagonal terms since $[h^{\text{s}}]_{+-(-+)}=\langle\psi_{+(-)}(0)|h|\psi_{-(+)}(0)
\rangle\sim\exp[-L/\xi]\approx 0$. Then, the non-Hermicity of 
chiral boundary states occurs only in diagonal terms, say 
$i\kappa_{0}\sigma_0$ and $i\kappa\sigma_3$. To realize Hermitian
chiral boundary states, the key is to find generic ways of removing 
$\kappa_0$ and $\kappa$ terms.
\par 

It is useful to understand the physics of the two non-Hermitian terms.
$i\kappa_{0}\sigma_0$ describes a loss (gain) of chiral boundary states 
for $\kappa_0<0$ ($\kappa_0>0$). Eigenstates of $h^{\text{s}}(p_1)$ 
in open systems are exponentially localized at ``boundaries'' of edges. 
The role of $\kappa\neq 0$ on chiral states can be seen by replacing 
$p_1$ by $(p_1+i\kappa/t)$ in the Bloch phase factor of $e^{ip_1 x}$. 
Such phenomena are termed as non-Hermitian skin effect~\cite{yao_prl_2018} 
and lead to higher-order skin-topological modes~\cite{chlee_prl_2019}. 
Accordingly, chiral boundary states of non-Hermitian topological 
insulators will be Hermitian when they neither have loss/gain nor exhibit 
the non-Hermitian skin effect in their propagating directions.
Below, three simple but instructive examples of Hermitian boundary 
states in non-Hermitian topological insulators are given.
\par

\emph{CIs.$-$}The first example is a non-Hermitian CI on a square 
lattice with lattice constant $a=1$, whose Bloch Hamiltonian reads
\begin{equation}
\begin{gathered}
h_1(\bm{k})=t\sum_{\mu=1,2}\sin k_\mu\sigma_\mu+\left(m+t\sum_{\mu=1,2}\cos k_\mu+i\kappa_3\right)\sigma_3
\end{gathered}\label{eq2}
\end{equation}
with $\sigma_{1,2,3}$ acting on the pseudospin spaces of two bulk bands. 
The Hermitian part of Eq.~\eqref{eq2} is the Qi-Wu-Zhang model.  
$m$ is the Dirac mass that controls the band inversion of bulk states 
\cite{qi_prb_2006}. Equation~\eqref{eq2} supports chiral edge states 
winding around the sample edge for $0<|m|<2t$. 
The non-Hermitian potential $i\kappa_3\sigma_3$ does not cause 
non-Hermitian skin effect in the $x-y$ plane such that the chiral edge 
states propagate along the sample edge without any localization. 
Equation~\eqref{eq2} has a particle-hole symmetry (PHS) because of 
$U_\mathcal{C}h^{T}_1(-\bm{k})U^{-1}_\mathcal{C}=-h_1(\bm{k})$ for a 
unitary operator $U_\mathcal{C}=\sigma_1$. This model belongs to class 
D of non-Hermitian Altland-Zirnbauer (AZ) symmetry classification~
\cite{kawabata_prx_2019}. 
\par

The effective low-energy Hamiltonian $h_1(p_1,p_2)$ of the CI
in the continuum limit can be obtained by expanding Eq.~\eqref{eq2} 
around $\Gamma$ point with $\bm{k}=\bm{p}+(0,0)$ 
($|\bm{p}|\ll 1$). For an infinite long strip of width $L$ with an 
open boundary condition (OBC) in the $y$ direction such that $p_1$ 
is a good quantum number and $p_2=-i\partial_y$, the eigenfunctions 
of $h_1(p_1,-i\partial_y)$ correspond to the edge states. 
Due to PHS, eigenenergies of $h_1(p_1,-i\partial_y)$ come in pairs 
of $\{\epsilon(p_1),-\epsilon(-p_1)\}$. Previous works show that 
$h_1(p_1,-i\partial_y)$ at $p_1=0$ has a vanishingly small energy 
band gap for $L\gg \xi$~\cite{mkonig_jps_2008,bzhou_prl_2008}. 
Then, we expect the existence of special edge states $\psi_{\pm}(p_1=0,y)$ 
with $\epsilon=0$. 
\par

\begin{figure}[htbp]
\centering
\includegraphics[width=0.45\textwidth]{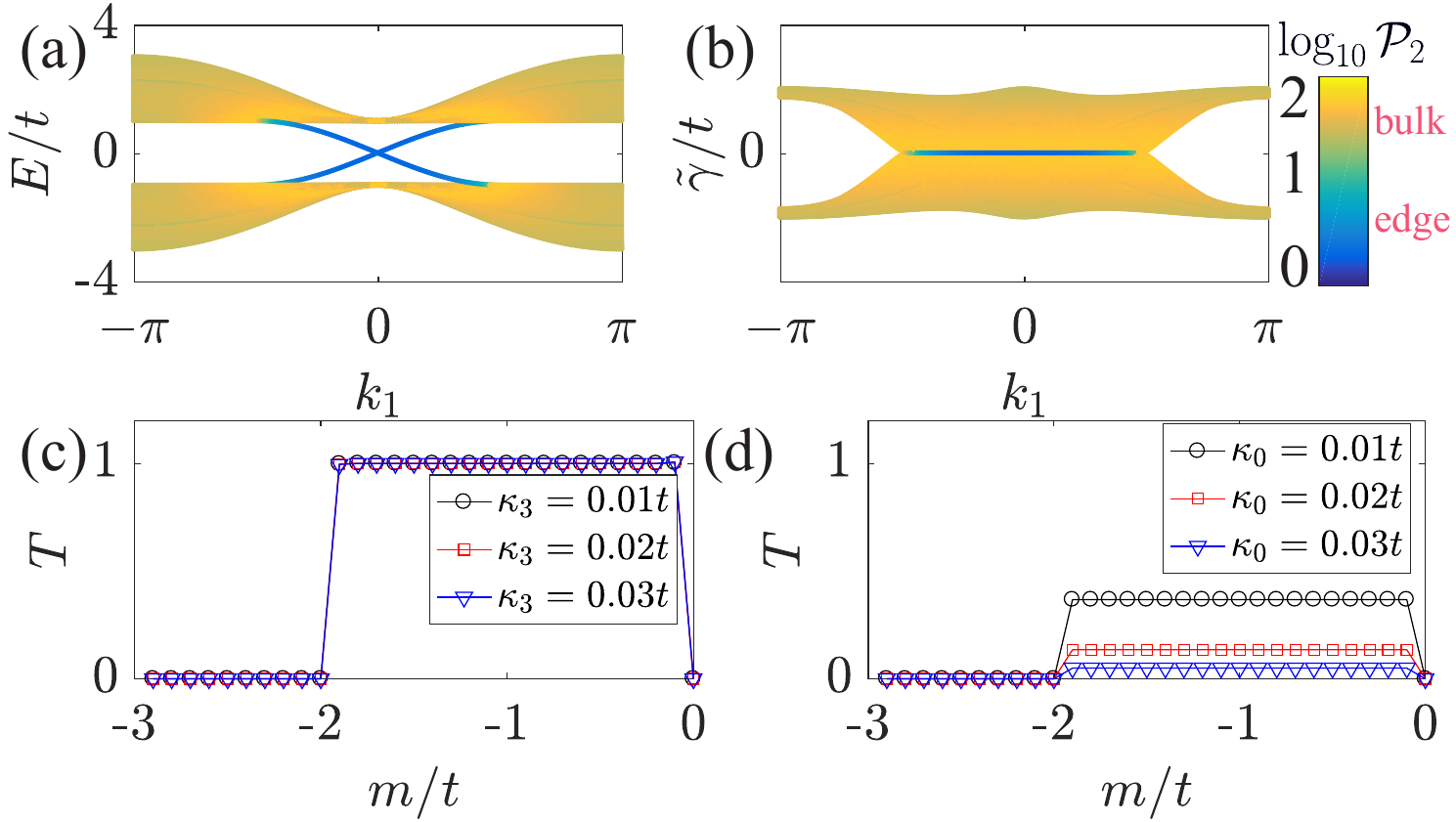}
\caption{(a) Real part of spectrum $E(k_1)/t$ of model~\eqref{eq2} 
for $L=40$ and $m=-t$. OBC is applied in the $y$ direction. Colors 
encode $\log_{10}\mathcal{P}_2$. (b) Imaginary part of spectrum 
$\tilde{\gamma}(k_1)/t$ for the same parameters used in (a). 
(c) Transmission coefficient $T$ as a function of $m/t$ for various $\kappa_3$. 
(d) $T$ v.s. $m/t$ with a global loss $-i\kappa_0 \sigma_0$.}
\label{fig1}
\end{figure}

To find out $\psi_{\pm}$, we solve the equation $h_1(p_1=0,-i\partial_y)\psi_{\pm}=0$ for one specific $m=-t$ and $0<\kappa_3\ll t$ and obtain the zero-energy edge states~\cite{supp}: $\psi_{\pm}=2\sqrt{2}e^{-L/2\pm y}\sinh[(i-\kappa_3/t)(y\mp L/2)]\sigma_2\phi_{\mp}$. Here, $\psi_{\pm}$ are localized at $y=\pm L/2$, and $\sigma_1\phi_{\pm}=\pm\phi_{\pm}$. The effective Hamiltonian of chiral edge states can be obtained by projecting Hamiltonian $h_1(p_1,-i\partial_y)$ into the space spanned by zero-energy edge states $\psi_{\pm}$, i.e., $[h^{\text{s}}_1]_{\alpha\beta}=\langle \psi_{\alpha}|h_1(p_1,-i\partial_y)|\psi_{\beta}\rangle$ with $\alpha,\beta=\pm$~\cite{mkonig_jps_2008,bzhou_prl_2008,jlinder_prb_2009,supp}: 
\begin{equation}
\begin{gathered}
h^{\text{s}}_1(p_1)=tp_1\sigma_3+O(p^2_1).
\end{gathered}\label{eq8}
\end{equation} 
Clearly, chiral edge states of Eq.~\eqref{eq2} with $m=-t$ are Hermitian. The non-Hermitian parameter $\kappa_3$ is encoded in $\psi_{\pm}$ and does not break the Hermicity of chiral edge states. 
\par

To verify the Hermicity of chiral edge states of Eq.~\eqref{eq2}, 
we compute its energy spectra $\epsilon(k_1)=E+i\tilde{\gamma}$ of a 
semi-infinite strip of width $L=40$ through Kwant package~\cite{kwant} 
and Scipy library~\cite{scipy} and plot $\epsilon(k_1)$ 
in Figs.~\ref{fig1}(a,b). Colors in Figs.~\ref{fig1}(a,b) encode the 
common logarithmic of participation ratios $\mathcal{P}_2$ of a right 
eigenstate $\psi_\epsilon$, defined as $\mathcal{P}_2(\epsilon)=
\langle(\sum_{\bm{i}}|\psi_\epsilon(\bm{i})|^4)^{-1}\rangle$, where 
$\psi_\epsilon(\bm{i})$ is the wave function amplitude at site $\bm{i}$. 
$\psi_\epsilon$ satisfies $h|\psi(\epsilon)\rangle=\epsilon|\psi(
\epsilon)\rangle$ and $\langle \psi_\epsilon|\psi_\epsilon\rangle=1$. 
$\mathcal{P}_2$ measures how many sites a state occupy from which one 
can easily identify bulk and edge states~\cite{cwang_prl_2015}. 
As shown in Figs.~\ref{fig1}(a,b), chiral edge states (the blue) of 
model~\eqref{eq2} are Hermitian with $\tilde{\gamma}=0$, while
bulk states (the yellow) are non-Hermitian.
\par

Noticeably, Eq.~\eqref{eq8} fails for a narrow width, say $L\sim\xi$, 
where edge states on opposite sites couple each other and open a gap 
in the complex energy plane~\cite{bzhou_prl_2008}. In this case, 
chiral edge states are non-Hermitian. Besides, we find $|E(p_1=0)|$ 
and $|\tilde{\gamma}(p_1=0)|$ decay in an exponentially law of $L/\xi$. 
In this work, we only focus on the large size limit $L\gg\xi$ where 
Hermitian chiral edge states are allowed.
\par

While quantized transport in the quantum Hall effect is one of 
the fundamental properties of Hermitian CIs, previous works show 
that Hall conductances of non-Hermitian CIs are non-quantized~\cite{philip_prb_2018,chen_prb_2018,groenendijk_prr_2021,yi_arXiv_2021}. 
The reason for the absence of quantized Hall conductances is that edge 
states of those CIs are non-Hermitian with certain degrees of loss. 
Naturally, for the Hermitian edge states predicted here, we expect 
perfectly quantized transport. To see it, we apply Eq.~\eqref{eq2} 
to a Hall bar of size $L\times L$ with two semi-infinite Hermitian 
leads at two ends along the $x$ direction and compute the 
transmission coefficient $T(\epsilon)$ by using B\"{u}ttiker's 
approach combined with the non-equilibrium Green function method, 
where a virtual lead of zero net particle flow is introduced to 
model the incoherent scatterings~\cite{datta}.
\par

\begin{figure}[htbp]
\centering
\includegraphics[width=0.45\textwidth]{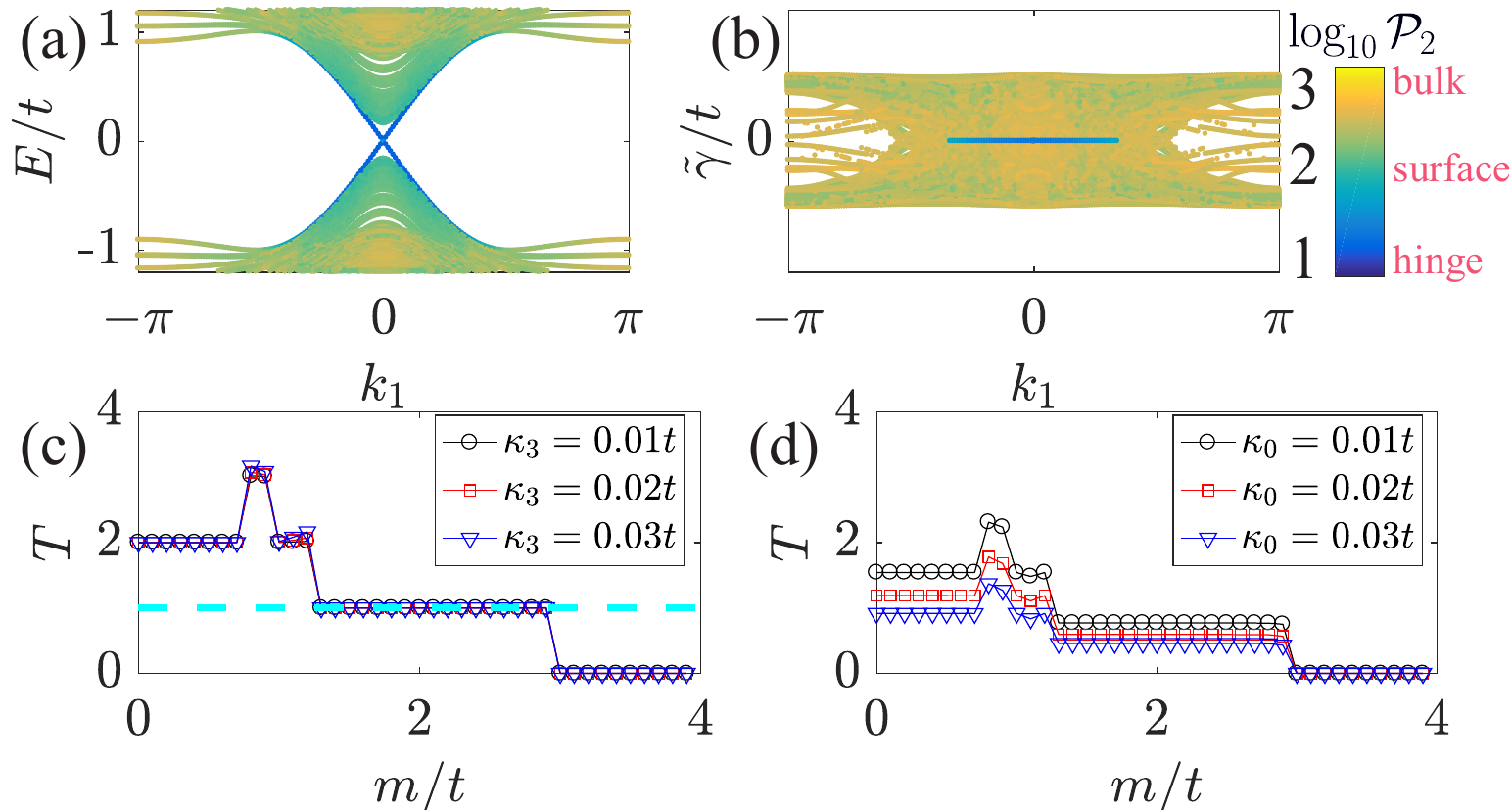}
\caption{(a) $E(k_1)/t$ of the non-Hermitian CI for $L=12$, $m=2t$, 
and $b=0.2t$. (b) $\tilde{\gamma}(k_1)/t$ for the same parameters of (a). 
Colors encode $\log_{10}\mathcal{P}_2$. 
(c) $T(0.02t)$ as a function of $m/t$ for various $\kappa_3$. 
(d) $T(0.02t)$ v.s. $m/t$ for a global loss $-i\kappa_0 \Gamma_0$. 
The cyan line locates $T=1$.}
\label{fig2}
\end{figure}

Figure~\ref{fig1}(c) displays some representative results with 
$\epsilon=0.02t$, $L=50$, and various $\kappa_3$. 
Here, $T(0.02t)$ is exactly quantized at 1 for $-2<m/t<0$, irrelevant 
to the values of $\kappa_3$.  For a comparison, we also plot $T(0.02t)$ 
for a different non-Hermitian CI, whose Hermitian part is the same 
as that of Eq.~\eqref{eq2} but the non-Hermicity is a global loss of 
$-i\kappa_0 \sigma_0$ ($\kappa_0>0$), see Fig.~\ref{fig1}(d),
where chiral edge states are non-Hermitian, and the transmission coefficient 
is non-quantized, qualitatively consistent with previous 
studies~\cite{philip_prb_2018,chen_prb_2018,groenendijk_prr_2021,yi_arXiv_2021}. 
\par

\emph{3DSOTI.$-$}The second example is a non-Hermitian 3DSOTI in a cubic 
lattice of $a=1$, which belongs to class A~\cite{cwang_prr_2020},  
\begin{equation}
\begin{gathered}
h_2(\bm{k})=t\sum_{\mu=1,2,3}\sin k_\mu\Gamma_\mu+\left(m-t\sum_{\mu=1,2,3}\cos k_\mu\right)
\Gamma_4\\
+(b+i\kappa_3)\Gamma_{13}.
\end{gathered}\label{eq9}
\end{equation}
Here, $\Gamma_{\mu=0,1,2,3,4,5}=(\sigma_0\otimes\sigma_0,\sigma_3\otimes
\sigma_1,\sigma_0\otimes\sigma_3,\sigma_1\otimes\sigma_1,\sigma_2\otimes
\sigma_1,\sigma_0\otimes\sigma_2)$ are the four-by-four identity and 
one particular choice of gamma matrices that satisfy $\{\Gamma_{\mu},
\Gamma_{\nu}\}=2\delta_{\mu,\nu}\Gamma_0$ and $\Gamma_{\mu\nu}=
[\Gamma_\mu,\Gamma_\nu]/(2i)$. $m$ and $b$ are the masses that control 
the band inversions of bulk states and surface states, respectively. 
For $0<b<1$ and $1+b<m<3-b$, the Hermitian part of model~\eqref{eq9} 
is a reflection-symmetric 3DSOTI with the reflection plane of $y=0$. 
If model~\eqref{eq9} is in a cubic lattice of size $L\times \sqrt{2}L\times \sqrt{2} L$ 
with OBCs on surfaces perpendicular to $(100),(01\bar{1}),(011)$, chiral hinge states 
will appear at the hinges when two surfaces meet at the reflection plane, e.g., 
$y=0,z=\pm L$.
\par

We use the same approach in Ref.~\cite{cwang_prb_2021} to derive 
the effective Hamiltonian of topological surface states of model~
\eqref{eq9} by replacing $p_2$ by $-i\partial_y$ around $\Gamma$ point:
\begin{equation}
\begin{gathered}
h^{\text{surface}}_2(\bm{p}_{\parallel}=(p_1,p_3))=
\begin{bmatrix}
h^{+}_2(\bm{p}_{\parallel}) & 0 \\
0 & h^{-}_2(\bm{p}_\parallel)
\end{bmatrix}
\end{gathered}\label{eq10}
\end{equation}
with $h^{\pm}_2(\bm{p}_{\parallel})=\pm[tp_1\sigma_1+tp_3\sigma_2-(b+i\kappa_3\pm t' p^2_{\parallel}/2)\sigma_2]$, 
see Supplemental Materials~\cite{supp}. 
Here, $m=2t$, and $\sigma_{1,2,3}$ act on the surface states of 
$\bm{p}_{\parallel}=(0,0)$. $t'$ decays exponentially with the 
distance of two surfaces, while $t'=-t/2$ if they meet at the 
reflection plane of $y=0$.
\par 

\begin{figure}[htbp]
\centering
\includegraphics[width=0.45\textwidth]{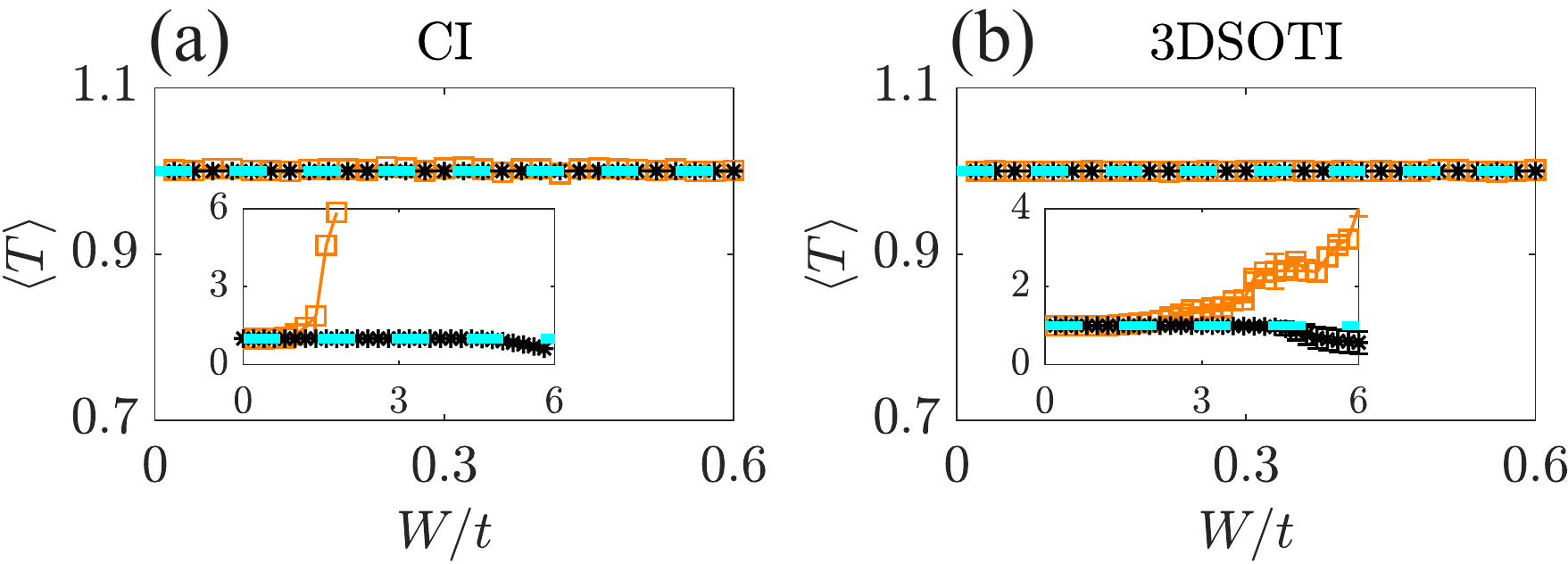}
\caption{(a) $\langle T(0.02t)\rangle$ as a function of $W/t$ for  
Hermitian (black stars) and non-Hermitian (orange squares) 
CIs with $m=-t$ and $L=50$. (b) Same as (a) for Hermitian (black 
stars) and non-Hermitian (orange squares) 3DSOTIs with $m=2t$, 
$b=0.2t$, and $L=12$. The cyan lines locate $\langle T(0.02t)\rangle=1$. 
Insets are for large-disorder regimes.}
\label{fig3}
\end{figure}

Equation~\eqref{eq10} indicates that the surface Hamiltonian of 3DSOTI act 
as a non-Hermitian CI with PHS. Band inversion of surface 
states, as well as the emergence of the hinge states, happens at 
$h^{+}_2(\bm{p}_\parallel)$ ($h^{-}_2(\bm{p}_\parallel)$) if 
$b>0$ ($b<0$). Since $h^{\pm}_2(\bm{p}_\parallel)$ does not suffer from 
the non-Hermitian skin effect, we expect its chiral hinge states 
are Hermitian. 
\par

Figures.~\ref{fig2}(a,b) display $E$ and $\tilde{\gamma}$ of Eq.~\eqref{eq9} 
with $L=12$, $m=2t$, $b=0.2t$, and OBCs on $(01\bar{1}),(011)$. We find 
chiral hinge states appear in the gap of surface states, 
where $\tilde{\gamma}=0$. Consequently, those chiral hinge states 
are Hermitian with $T=1$, see Fig.~\ref{fig2}(c). 
While, if we replace $i\kappa_3\Gamma_{13}$ by a global loss 
$-i\kappa_0 \Gamma_0$, the chiral hinge states are non-Hermitian 
without quantized $T$, see Fig.~\ref{fig2}(d).
\par

Equation~\eqref{eq9} also supports Hermitian helical hinge states 
with valley-momentum locking when $0<m<1-b$~\cite{cwang_prr_2020}. 
$T$ for Hermitian helical hinge states is quantized (to 2) for 
$i\kappa_3\Gamma_{13}$, irrelevant to the value of $\kappa_3$, 
see Fig.~\ref{fig2}(c). However, these hinge states will loss 
the quantized transport in the presence of disorders due to the 
facilitation of inter-valley scattering and mixing.   
\par

\emph{Robustness against disorders.$-$}To fully establish these 
Hermitian edge/hinge states as genuine states of matter, their 
robustness against disorders should be tested. Here, white-noise 
on-site random complex potentials, 
$V_1=\sum_{\bm{i}}c^\dagger_{\bm{i}}(v^r_{\bm{i}}\sigma_0+iv^i_{\bm{i}}
\sigma_3) c_{\bm{i}}$ and $V_2=\sum_{\bm{i}}c^\dagger_{\bm{i}}
(v^r_{\bm{i}}\Gamma_0+iv^i_{\bm{i}}\Gamma_{31}) c_{\bm{i}}$, are added  
to lattice models of Eqs.~\eqref{eq2} and \eqref{eq9}, respectively. 
$c^\dagger_{\bm{i}}$ and $c_{\bm{i}}$ are the creation and the 
annihilation operators on site $\bm{i}$, and $v^{r,i}_{\bm{i}}$ 
distribute uniformly in $[-W_r/2,W_r/2]$ and $[-W_i/2,W_i/2]$, where  
real numbers $W_{r,i}$ measure the strength of randomness.  
\par

Figures~\ref{fig3}(a,b) depict the ensemble-averaged $\langle T
(W=W_r=W_i)\rangle$ for disordered non-Hermitian CIs  ($m=t$) and 
3DSOTIs ($m=2t$ and $b=0.2t$) of $\epsilon=0.02t$, respectively. 
For a comparison, $\langle T\rangle$ for their Hermitian counterparts 
($\kappa_3=W_i=0$) are also computed. Evidently, quantized plateaus 
of $\langle T\rangle$ persist until a finite disorder $W_c$ (depends on 
model parameters) for both non-Hermitian CI and 3DSOTI. Such plateaus 
are size-independent (not shown here) with extremely small fluctuations 
of $T$, which should be convincing supports of the robustness of Hermitian 
chiral edge/hinge states. However, such chiral edge/hinge states loss their Hermicity
if disorders lead to an effective gain/loss, see Supplemental Materials~\cite{supp}.
\par

\begin{figure}[htbp]
\centering
\includegraphics[width=0.45\textwidth]{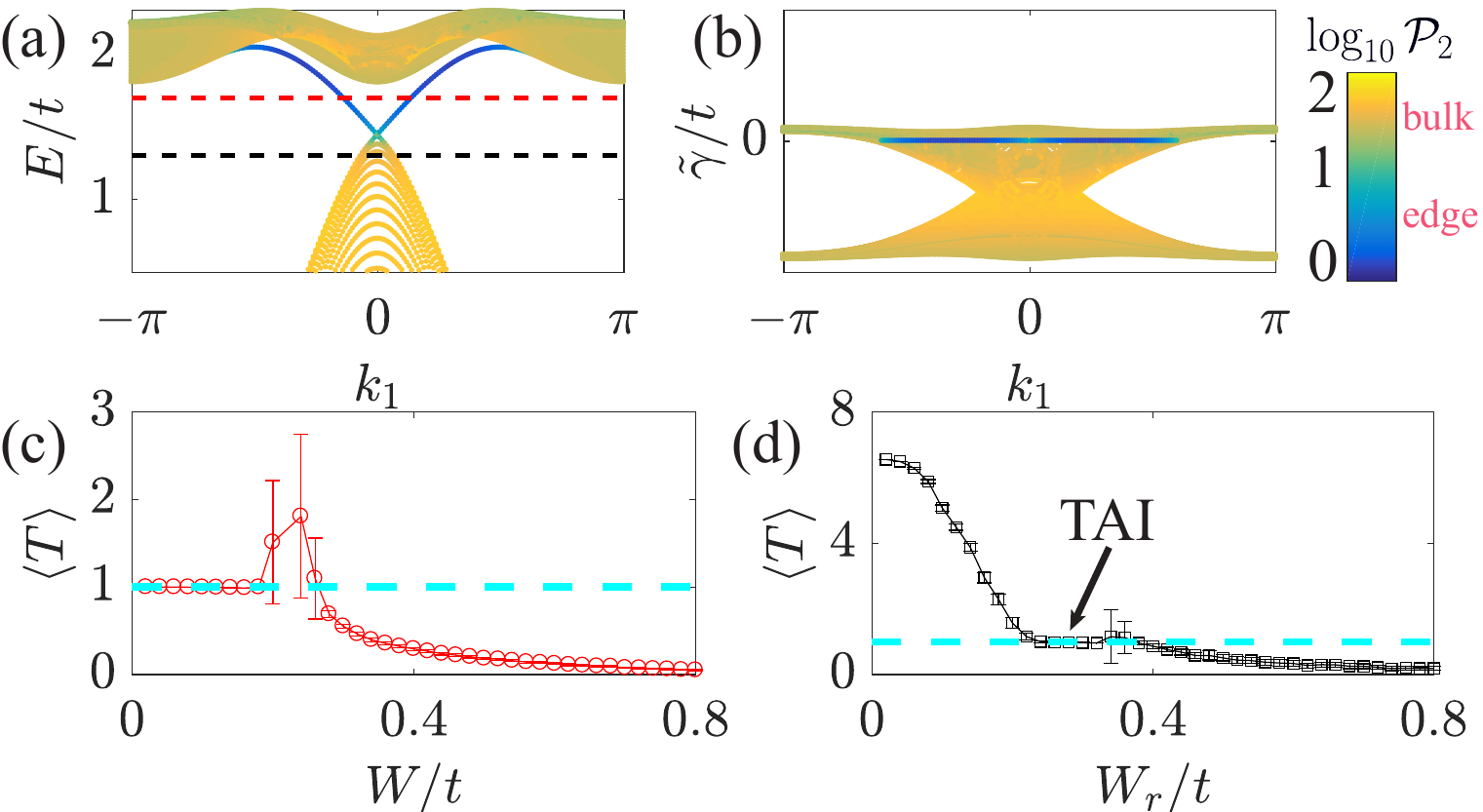}
\caption{(a,b) Real (a) and imaginary (b) parts of spectrum 
$\epsilon(k_1)$ of model~\eqref{eq12}. Colors encode 
$\log_{10}\mathcal{P}_2$. The dash lines in (a) denote 
the Fermi energies for $\langle T\rangle$ in (c,d). 
(c,d) $\langle T\rangle$ v.s. $W/t$ and $W_r/t$ for $E=1.3t$ (c) 
and $E=1.5t$ (d). Here, $m=1.8t$, $t_0=-0.8t$, $\kappa_3=0.01t$, 
$\kappa_0=-1.2\kappa_3$, and $L=100$. The cyan lines denote 
$\langle T\rangle=1$.}
\label{fig4}
\end{figure}

Interestingly, Hermitian and non-Hermitian systems have qualitatively 
different transport behaviors for $W>W_c$. In Hermitian systems, 
$\langle T\rangle<1$ for $W>W_c$ where backward scatterings lead to 
a non-zero reflection probability; while in non-Hermitian systems, 
$\langle T\rangle$ could be larger than 1 since the transmission 
amplitude can be amplified by the non-Hermicity.
\par

\emph{TAIs.$-$}Saliently, one can also construct Hermitian edge 
states without PHS, where a global loss/gain is required. 
This can be seen by introducing a new term $t_0\sum_{\mu=1,2}\cos 
k_\mu \sigma_0$, which breaks PHS, as well as a compensated loss/gain 
$i\kappa_0 \sigma_0$ in the non-Hermitian CI [Eq.~\eqref{eq2}]:
\begin{equation}
\begin{gathered}
h_3(\bm{k})=h_1(\bm{k})+\left(t_0\sum_{\mu=1,2}\cos k_\mu+
i\kappa_0\right)\sigma_0.
\end{gathered}\label{eq12}
\end{equation}
Equation~\eqref{eq12} belongs to class A. Near $p_1=0$, the effective Hamiltonian for chiral edge states reads 
(to the order of $p_1$, and $\kappa_0,\kappa_3\ll t$)~\cite{bzhou_prl_2008,supp}
\begin{equation}
\begin{gathered}
h^{\text{s}}_3(p_1)=[-mt_0/t+i(\kappa_0-t_0\kappa_3/t)]\sigma_0+
\sqrt{t^2-t^2_0}p_1\sigma_3.
\end{gathered}\label{eq11}
\end{equation}
Hence, chiral edge states are Hermitian when $\kappa_0=t_0\kappa_3/t$ 
and $|t_0|<|t|$, confirmed in the energy spectrum shown in Figs.~\ref{fig4}(a,b) for 
$m=1.8t$, $t_0=-0.8t$, $\kappa_3=0.01t$, and $\kappa_0=-0.8\kappa_3$. 
Such chiral edge states are robust against the random potentials 
$V_1$ until $W_c=0.25t$, supported by the quantized ensemble-average 
transmission coefficient $\langle T(1.5t)\rangle$ in Fig.~\ref{fig4}(c).
\par

In Hermitian systems, TAI, a disorder-induced topological insulator, 
could emerge in disordered PHS-broken CIs, i.e., the Hermitian part 
of Eq.~\eqref{eq12} subject to the random potential $V_1$ with 
$W_i=0$~\cite{jli_prl_2009,cwgroth_prl_2009}. TAIs are featured 
by quantized transmission coefficients at finite disorders when 
it is topological trivial in crystals. Naturally, one may ask 
whether there is a non-Hermitian TAI characterized by quantized 
transport only at finite disorders. Numerical evidence to the 
existence of non-Hermitian TAIs is shown in Fig.~\ref{fig4}(d): 
There is a plateau of $\langle T(1.3t)\rangle=1$ at finite disorders 
$W_r/t\in[0.25,0.3]$ and $W_i=0$, when the system is a 
topologically-trivial metal at weak disorders $W_r/t\in[0,0.25]$ 
and becomes a topologically-trivial insulator at strong disorders $W_r>0.3t$. 
\par

\emph{Discussions.$-$}Hermitian chiral boundary states in non-Hermitian 
topological insulators reported here are different from real spectra in
non-Hermitian systems with PT-symmetry~\cite{cmbender_prl_1998,hyang_prl_2018,kkawabata_prr_2020}. 
There exist exceptional points in PT-symmetric systems that separate real 
spectra in PT-symmetry preserved region from complex spectra in 
PT-symmetry broken region. Such exceptional points do not 
occur in the Hermitian chiral boundary states. Hermicity of the 
boundary states is insensitive to smooth changes of material parameters 
and cannot be destroyed unless the system undergoes topological phase transitions.
\par

Electronic CIs and 3DSOTIs have been predicted and experimentally 
observed in many materials such as HgTe~\cite{bernevig_science_2006,
konig_science_2007} and Bi$_{2-x}$Sm$_x$Se$_3$~\cite{cyue_natphys_2019}. 
However, it is difficult to satisfy the condition of $\kappa_0=t_0
\kappa_3/t$ even one can manipulate lifetimes of electrons in 
different bands, thus $i(\kappa_0 \sigma_0+\kappa_3\sigma_3)$ in CIs. 
In contrast, recent progress in cold-atom, photonic, and optomechanical 
systems~\cite{gbarontini_prl_2013,hxu_nature_2016,wchen_nature_2017} 
allows us to control the gain and loss of excitation states in those 
systems. For example, a topological insulator laser can possess Hermitian chiral
edge states in two dimensions, see Supplemental Materials~\cite{supp}.
\par

\emph{Conclusion.$-$}In conclusion, we showed that it is possible to make
the chiral edge/hinge states of non-Hermitian topological insulator passive
without gain/loss and immune to non-Hermitian skin effect. Consequently,
these chiral edge/hinge states become Hermitian and give rise to quantized 
edge/hinge transport in non-Hermitian topological insulators. Remarkably, their 
Hermicity does not rely on any symmetry, e.g., one can have Hermitian 
chiral edge/hinge states in both class D [respect to particle-hole symmetry] 
and class A [without symmetry] in AZ symmetry classification.
\par

\begin{acknowledgments}

This work is supported by the National Natural Science Foundation of China 
(Grants No.~11774296, No.~11704061,and No.~11974296),
the National Key Research and Development Program of China 2020YFA0309600, 
and Hong Kong RGC (Grants Nos.~16301518 16301619, and 16302321).

\end{acknowledgments}

\newpage

\section{Supplemental Materials: Hermitian chiral boundary states in non-Hermitian topological insulators}

These Supplemental Materials contain the following information: (S1) Derivations of Eqs.~(3,5,7) in the main text. (S2) Disorder effects on the Hermicity of chiral edge states. (S3) Coupled lasers in a honeycomb lattice display the predicted Hermitian chiral edge states. 

\subsection*{S1: Derivations of Eqs.~(3,5,7) in the main text}

In this section, we show how to derive Eqs.~(3,5,7) in the main text.\par

\quad\par
\emph{Equation~(3): Chern insulators}.\par
\quad\par

Near the $\Gamma$ point ($\bm{k}=(0,0)$), the low-energy effective Hamiltonian of the non-Hermitian Chern insulator (CI) reads
\begin{equation}
\begin{gathered}
h_1(\bm{p})=tp_1\sigma_1+tp_2\sigma_2+\left(\tilde{m}+i\kappa_3 -\dfrac{t}{2}p^2_1-\dfrac{t}{2}p^2_2\right)\sigma_3.
\end{gathered}\label{eq_h_1_1}
\end{equation}
with $\tilde{m}=m+2t$. For one specific parameter of $m=-t$ and $0<\kappa_3\ll t$, we have
\begin{equation}
\begin{gathered}
h_1(\bm{p})=t\left[p_1\sigma_1+p_2\sigma_2+\left(1+i\tilde{\kappa}_3-\dfrac{1}{2}p^2_1-\dfrac{1}{2}p^2_2\right)\sigma_3\right]
\end{gathered}\label{eq_h_1_2}
\end{equation}
with $\tilde{\kappa}_3=\kappa_3/t$. Periodic boundary condition (PBC) is applied in the $x$-direction, and open boundary condition (OBC) is applied in the $y$-direction such that $p_2\to -i\partial_y$ and $-L/2<y<L/2$. We split Eq.~\eqref{eq_h_1_2} into two parts:
\begin{equation}
\begin{gathered}
h^1_1(-i\partial_y)=t\left[-i\partial_y\sigma_2+\left(1+i\tilde{\kappa}_3+\dfrac{1}{2}\partial_{yy}\right)\sigma_3\right]
\end{gathered}\label{eq_h_1_3}
\end{equation}
and
\begin{equation}
\begin{gathered}
h^2_1(p_1)=t\left(p_1\sigma_1-\dfrac{1}{2}p^2_1\sigma_3\right).
\end{gathered}\label{eq_h_1_4}
\end{equation}
Due to particle-hole symmetry, eigenenergies of $h_1(p_1,-i\partial_y)$ come in pairs of $\{\epsilon(p_1),-\epsilon(-p_1)\}$. To find special edge state wavefunction $\psi(p_1=0,y)\equiv \psi(y)$ of $\epsilon=0$, we solve $h^1_1(-i\partial_y)\psi=0$ for solution of form $\psi(y)=i\sigma_2 \phi e^{\lambda_y}$:
\begin{equation}
\begin{gathered}
t\left[\lambda\phi+\left(1+i\tilde{\kappa}_3+\dfrac{\lambda^2}{2}\right)\sigma_1\phi\right]=0.
\end{gathered}\label{eq_h_1_5}
\end{equation}
$\phi$ should be the eigenvectors of $\sigma_1$, say $\sigma_1\phi_{\pm}=\pm\phi$. For $\phi_+$, $\lambda$ satisfies  
\begin{equation}
\begin{gathered}
\lambda+\left(1+i\tilde{\kappa}_3+\dfrac{\lambda^2}{2}\right)=0 \to \lambda_{1,2}= -1\pm \left( i-\tilde{\kappa}_3\right).
\end{gathered}\label{eq_h_1_6}
\end{equation}
For $\phi_-$, $\lambda$ is given by
\begin{equation}
\begin{gathered}
\lambda-\left(1+i\tilde{\kappa}_3+\dfrac{\lambda^2}{2}\right)=0 \to \lambda_{3,4}=1\pm(i-\tilde{\kappa}_3).
\end{gathered}\label{eq_h_1_7}
\end{equation}
Hence,
\begin{equation}
\begin{gathered}
\psi(y)=i\sigma_2\left[\left(c_1 e^{\lambda_1 y}+c_2 e^{\lambda_2 y}\right)\phi_+ + \left(d_1 e^{\lambda_3 y}+d_2 e^{\lambda_4 y}\right)\phi_-\right]
\end{gathered}\label{eq_h_1_8}
\end{equation}
with $\text{Re}[\lambda_{3,4}]=-\text{Re}[\lambda_{1,2}]>0$. For edge states localized at $y=-L/2$, we requires $d_1=d_2=0$ and $\psi(y=-L/2)=0$. Therefore, the solution of edge states localized at $y=-L/2$ reads
\begin{equation}
\begin{gathered}
\psi_{-}(y)=i2ce^{-L/2-y}\sinh \left[ (i-\tilde{\kappa}_3)(L/2+y) \right]\sigma_2\phi_+ 
\end{gathered}\label{eq_h_1_9}
\end{equation} 
with $c$ being a normalized constant. To normalize $\psi_{-}(y)$ yields $|c|^2=2$ if $\tilde{\kappa}_3\ll 1$ and $L\gg 1$, and we choose $c=-i\sqrt{2}$ such that
\begin{equation}
\begin{gathered}
\psi_{-}(y)=2\sqrt{2}e^{-L/2-y}\sinh \left[ (i-\tilde{\kappa}_3)(L/2+y) \right]\sigma_2\phi_+.
\end{gathered}\label{eq_h_1_10}
\end{equation}
Following the same approach, we find the edge states localized on the other side $y=L/2$,  
\begin{equation}
\begin{gathered}
\psi_{+}(y)=2\sqrt{2} e^{-L/2+y}\sinh[(i-\tilde{\kappa}_3)(y-L/2)]\sigma_2 \phi_-.
\end{gathered}\label{eq_h_1_11}
\end{equation}
The effective Hamiltonian for the zero-energy chiral edge states is obtained by projecting bulk Hamiltonian Eq.~\eqref{eq_h_1_3} onto space spanned by $\psi_{\pm}$, i.e., $[h^s_1]_{\alpha\beta}=\langle\psi_{\alpha}|h_1(p_1,-i\partial_y)|\psi_{\beta}\rangle$ with $\alpha,\beta=\pm$~\cite{s_mkonig_jps_2008}. Following this approach, we obtain $h^s_1=tp_1\sigma_3+O(p^2_1)$, which is Eq.~(3) in the main text.\par

\quad\par
\emph{Equation~(5): Three-dimensional second-order topological insulators}.\par
\quad\par

Let us expand Eq.~(4) in the main text near the Gamma point, say $\bm{k}=(0,0,0)+\bm{p}$ with $\bm{p}\ll 1$:
\begin{equation}
\begin{gathered}
h_{2}(\bm{p})=tp_1\Gamma_1+tp_2\Gamma_2+tp_3\Gamma_3\\
+\left[\tilde{m}+\dfrac{t}{2}\left(p^2_1+p^2_2+p^2_3\right)\right]\Gamma_4+(b+i\kappa_3)\Gamma_{31}
\end{gathered}\label{lh3dsoti_1_2}
\end{equation}
with $\tilde{m}=m-3t$. Then, we apply the OBC in the $y$ direction, i.e., $p_2\to -i\partial_y$ and $-L/2<y<L/2$, and divide Eq.~\eqref{lh3dsoti_1_2} into two parts:
\begin{equation}
\begin{gathered}
h^1_{2}(-i\partial_y)=-it\partial_y\Gamma_2+\left(\tilde{m}-\dfrac{t}{2}\partial_{yy}\right)\Gamma_4
\end{gathered}\label{lh3dsoti_1_3}
\end{equation}
and
\begin{equation}
\begin{gathered}
h^2_{2}(\bm{p}_{\parallel}=(p_1,p_3))=tp_1\Gamma_1+tp_3\Gamma_3+\dfrac{t}{2}\left(p^2_1+p^2_3\right)\Gamma_4\\
+(b+i\kappa_3)\Gamma_{31}.
\end{gathered}\label{lh3dsoti_1_4}
\end{equation}
We expect a zero-energy solution of Eq.~\eqref{lh3dsoti_1_3}: $h^1_{2}(-i\partial_y)\psi=0$. To obtain $\psi$, we apply a unitary transformation $\tilde{h}_{2}(\bm{p}_{\parallel},-i\partial_y)=Uh_{2}(\bm{p}_{\parallel},-i\partial_y) U^{-1}$ with $U=\exp[i\pi \sigma_1/4]\otimes\exp[-i\pi \sigma_1/4]$. Then, we solve $\tilde{h}^1_{2}(-i\partial_y)\tilde{\psi}=0$ where
\begin{widetext}
\begin{equation}
\begin{gathered}
\tilde{h}^1_{2}(-i\partial_y)=
\begin{bmatrix}
0 & t\partial_y+\tilde{m}-t\partial_{yy}/2 & 0 & 0 \\
-t\partial_y+\tilde{m}-t\partial_{yy}/2 & 0 & 0 & 0 \\
0 & 0 & 0 & t\partial_y-\tilde{m}+t\partial_{yy}/2 \\
0 & 0 & -t\partial_y-\tilde{m}+t\partial_{yy}/2 & 0
\end{bmatrix}
\end{gathered}\label{lh3dsoti_1_6}
\end{equation}
\end{widetext}
and $\tilde{\psi}=U\psi$. Consider a specific value $m=2t$. It is clear that $\tilde{\psi}$ have four solutions:
\[
\tilde{\psi}_1=\tilde{\phi}_1
\begin{bmatrix}
1 \\
0 \\
0 \\
0
\end{bmatrix},
\tilde{\psi}_2=\tilde{\phi}_2
\begin{bmatrix}
0 \\
1 \\
0 \\
0
\end{bmatrix},
\tilde{\psi}_3=\tilde{\phi}_3
\begin{bmatrix}
0 \\
0 \\
1 \\
0
\end{bmatrix},
\tilde{\psi}_4=\tilde{\phi}_4
\begin{bmatrix}
0 \\
0 \\
0 \\
1
\end{bmatrix}.
\]
Use the trial solution $\tilde{\phi}_{1,2,3,4}=e^{\lambda_{1,2,3,4}y}$. $\lambda_1$ satisfies the quadratic equation $\left(-t\partial_y-t-t\partial_{yy}/2\right)e^{\lambda_1 y}=0$. We find $\lambda_1=-1\pm i$ such that $\tilde{\phi}_1$ localizes at $y=-L/2$. From boundary condition $\tilde{\phi}_1(y=-L/2)=0$, we find $\tilde{\phi}_1(y)=i2ce^{-y}\sin[y+L/2]$ with $c$ being a normalized constant. Normalization condition gives $|c|^2=(2e^{-L})/[1+e^{-2L}(\cos[2L]-\sin[2L]-2)]$, and we choose $c=(-i\sqrt{2}e^{-L/2})/([1+e^{-2L}(\cos[2L]-\sin[2L]-2)]^{1/2})$. Then we have
\begin{equation}
\begin{gathered}
\tilde{\phi}_1(y)=\dfrac{2\sqrt{2}e^{-(y+L/2)}\sin[y+L/2]}{[1+e^{-2L}(\cos[2L]-\sin[2L]-2)]^{1/2}}.
\end{gathered}\label{lh3dsoti_1_7}
\end{equation}
Likewise, we obtain
\begin{equation}
\begin{gathered}
\tilde{\phi}_2(y)=\dfrac{2\sqrt{2}e^{y-L/2}\sin[y-L/2]}{[1+e^{-2L}(\cos[2L]-\sin[2L]-2)]^{1/2}},
\end{gathered}\label{lh3dsoti_1_8}
\end{equation}
and
\begin{equation}
\begin{gathered}
\tilde{\phi}_3(y)=\tilde{\phi}_2(y),~\tilde{\phi}_4(y)=\tilde{\phi}_1(y).
\end{gathered}\label{lh3dsoti_1_10}
\end{equation}
The surface Hamiltonian can be obtained by projecting the bulk Hamiltonian into the space spanned by the four specific surface states~\cite{cwang_prb_2021}
\begin{equation}
\begin{gathered}
h^{\text{surface}}_2(\bm{p}_\parallel) =\int\begin{bmatrix}
\tilde{\psi}^\dagger_1 \\
\tilde{\psi}^\dagger_2 \\
\tilde{\psi}^\dagger_3 \\
\tilde{\psi}^\dagger_4 \\
\end{bmatrix}
\left( Uh_2(\bm{p}_\parallel,-i\partial_y) U^{-1} \right)
\begin{bmatrix}
\tilde{\psi}_1 & \tilde{\psi}_2 & \tilde{\psi}_3 & \tilde{\psi}_4
\end{bmatrix}dy\\
=tp_1\sigma_1\otimes\sigma_1+tp_3\sigma_2\otimes\sigma_1-(b+i\kappa_3)\sigma_3\otimes\sigma_0\\
-\dfrac{t'}{2}(p^2_1+p^2_3)\sigma_3\otimes\sigma_1.
\end{gathered}\label{lh3dsoti_1_13}
\end{equation}
Again, we perform a unitary transformation $U'=\sigma_0\otimes\exp[-\pi\sigma_3/4]$ to Eq.~\eqref{lh3dsoti_1_13} and write the effective Hamiltonian for the surface states in the following form:
\begin{equation}
\begin{gathered}
h^{\text{surf}}_{2}=
\begin{bmatrix}
h^+_2(\bm{p}_{\parallel}) & 0 \\
0 & h^-_2(\bm{p}_{\parallel})
\end{bmatrix}
\end{gathered}\label{lh3dsoti_1_16}
\end{equation}
with $h^\pm_2(\bm{p}_{\parallel})=\pm[tp_1\sigma_1+tp_3\sigma_2-(b+i\kappa_3\pm t'(p^2_1+p^2_3)/2)\sigma_3]$. Equation~\eqref{lh3dsoti_1_16} is Eq.~(5) in the main text. Here, we introduce a new quantity $t'$:
\begin{equation}
\begin{gathered}
t'=t\int^{L/2}_{-L/2}\tilde{\phi}^\ast_1(y)\tilde{\phi}_2(y)dy=t\dfrac{4e^{-L}(L\cos[L]-\sin[L])}{[1+e^{-2L}(\cos[2L]-\sin[2L]-2)]}
\end{gathered}\label{lh3dsoti_1_18}
\end{equation}
Equation~\eqref{lh3dsoti_1_18} vanishes for $L\gg 1$ since $t'$ decreases exponentially with $L$. On the other hand, at the hinges where two surfaces meet $L\ll 1$, $t'=-t/2$.
\par

\quad\par
\emph{Equation~(7): Topological Anderson insulators}.\par
\quad\par

We expand Eq.~(7) in the main text near the $\Gamma$ point to obtain the low-energy effective Hamiltonian:
\begin{equation}
\begin{gathered}
h_3(\bm{p})=tp_1\sigma_1+tp_2\sigma_2+\left[\tilde{m}+i\kappa_3-\dfrac{t}{2}\left(p^2_1+p^2_2\right)\right]\sigma_3\\
+\left[2t_0+i\kappa_0-\dfrac{t_0}{2}\left(p^2_1+p^2_2\right)\right]\sigma_0
\end{gathered}\label{eq_stai_1}
\end{equation}
with $\tilde{m}=m+2t$. We now solve the eigenfunction of the following Hamiltonian:
\begin{equation}
\begin{gathered}
\tilde{h}_3(\bm{p})=h_3(\bm{p})-\left(2t_0+i\kappa_0\right)\sigma_0=tp_1\sigma_1+tp_2\sigma_2\\
+\left[\tilde{m}+i\kappa_3-\dfrac{t}{2}\left(p^2_1+p^2_2\right)\right]\sigma_3-\dfrac{t_0}{2}\left(p^2_1+p^2_2\right)\sigma_0
\end{gathered}\label{eq_stai_2}
\end{equation}
in a finite strip of a width $L$ with the PBC applied in the $x$ direction and the OBC applied in the $y$ direction. Now, $p_1$ is still a good quantum number, and $p_2=-i\partial_y$, i.e.,
\begin{equation}
\begin{gathered}
\left\{
\begin{array}{c}
\left[ \tilde{m}+i\kappa_3-b_+(p^2_1-\partial_{yy}) \right]\psi_1+t(p_1-\partial_y)\psi_2=\epsilon\psi_1 \\
\quad \\
t(p_1+\partial_y)\psi_1-\left[ \tilde{m}+i\kappa_3- b_- \left(p^2_1-\partial_{yy}\right) \right]\psi_2 =\epsilon\psi_2
\end{array}
\right.
\end{gathered}\label{eq_stai_3}
\end{equation}
with $b_+=(t+t_0)/2$ and $b_-=(t-t_0)/2$. From Eq.~\eqref{eq_stai_3}, one can find the energy spectrum $\epsilon$ for $L\gg 1$ near $p_1=0$ by using the trial function $\psi_{1,2}=e^{\lambda y}$ [The details are illustrated in Ref.~\cite{bzhou_prl_2008}]: 
\begin{equation}
\begin{gathered}
\epsilon_{\pm}(p_1)=-\dfrac{t_0}{t}\left(\tilde{m}+i\kappa_3\right)\pm\sqrt{t^2-t^2_0}p_1
\end{gathered}\label{eq_stai_4}
\end{equation}  
providing $t^2>t^2_0$. Recall that $\epsilon_{\pm}(p_1)$ are the dispersions for Eq.~\eqref{eq_stai_2}, but not that of Eq.~\eqref{eq_stai_1}, where a constant complex energy $\left(2t_0+i\kappa_0\right)$ is shifted. Hence, the edge state spectrum of the original Hamiltonian $h_3(\bm{p})$ should be
\begin{equation}
\begin{gathered}
\epsilon_{\pm}(p_1)=\left[ -t_0 m/t+i(\kappa_0-t_0\kappa_3/t)\right]\pm\sqrt{t^2-t^2_0}p_1
\end{gathered}\label{eq_stai_5}
\end{equation}
We use the same approach for obtaining the effective Hamiltonian of chiral edge states of the non-Hermitian CI: $[h^s_3]_{\alpha\beta}=\langle\psi_\alpha|h_3(p_1,-i\partial_y)|\psi_\beta\rangle$ with $\alpha,\beta=\pm$ and $h_3(p_1,-i\partial_y)|\psi_\alpha\rangle=\epsilon_{\alpha}(p_1)|\psi_\alpha\rangle$:
\begin{equation}
\begin{gathered}
h^s_3(p_1)=\left[ -t_0 m/t+i(\kappa_0-t_0\kappa_3/t)\right]\sigma_0+\sqrt{t^2-t^2_0}p_1\sigma_3.
\end{gathered}\label{eq_stai_6}
\end{equation}
Equation~\eqref{eq_stai_6} is Eq.~(7) in the main text.
\par

\subsection*{S2: Disorder effects on the Hermicity of chiral edge states}

The keys for realizing Hermitian chiral edge states are: (i) chiral edge states have no gain and no loss; (ii) the chiral edge states can propagate along sample boundaries without suffering from the non-Hermitian skin effect [Otherwise, they become corner states]. Therefore, any disorders that lead to the gain/loss or the non-Hermitian skin effect on chiral edge states break their Hermicity. Within a simple self-consistent Born approximation, the effective Hamiltonian of a disordered system is $h_{\text{eff}}(\bm{k})=h_{0}(\bm{k})+\Sigma$ with $h_{0}(\bm{k})$ being the Hamiltonian of a clean system and $\Sigma$ being the disorder-induced self-energy. In the continuous limit, the self-energy in the first-order reads~\cite{book1}
\begin{equation}
\begin{gathered}
\Sigma^1=\dfrac{1}{V}\int U(\bm{x})d^{d}x=\langle U(\bm{x})\rangle.
\end{gathered}\label{eq2_10}
\end{equation}
Here, $U(\bm{x})$ describes the profile of random potentials, and $V$ is the volume of the system. Hence, the first-order self-energy shifts the complex-energy spectrum by a constant proportional to the disordered potential's spatial average. In the main text where $\langle U(\bm{x})\rangle=0,\langle U^2(\bm{x})\rangle=W^2/12$, this average is zero. Thus, disorders do not lead to gain/loss for chiral boundary states. However, for disorders that is $\langle U(\bm{x})\rangle=i\kappa$, a global gain ($\kappa>0$) or loss ($\kappa<0$) arise from the disorders. The Hermicity of the chiral edge states is lost.\par

\begin{figure}[htbp]
\centering
\includegraphics[width=0.4\textwidth]{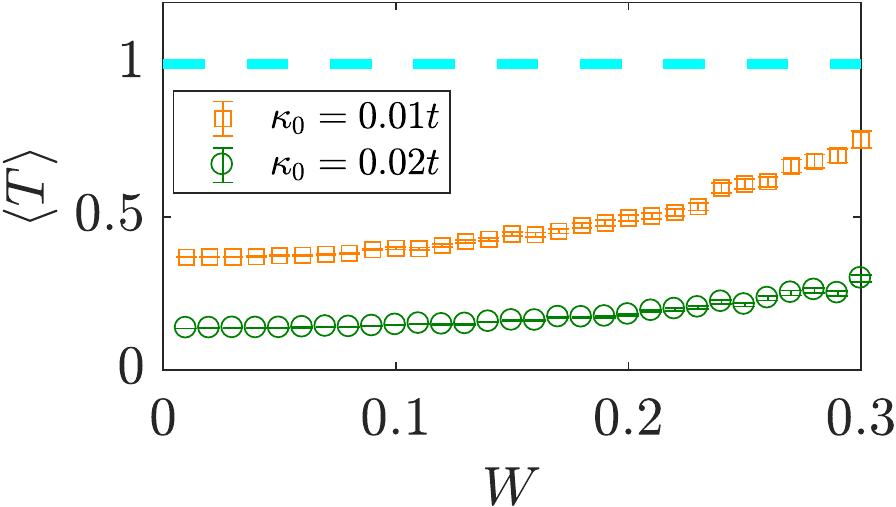}
\caption{Ensemble-averaged transmission coefficient $\langle T(0.02t)\rangle$ as a function of disorder $W$ for $\kappa=0.01,0.02$ for non-Hermitian Chern insulators [Parameters are the same as those for Fig.~3 in the main text]. The cyan line locate $\langle T(0.02t)\rangle=1$.}
\label{fig_reply_2}
\end{figure}

To illustrate the above point, we consider the following disorders
\begin{equation}
\begin{gathered}
V_{\text{broken}}=\sum_{\bm{i}} c^\dagger_{\bm{i}} \left(i\tilde{v}^i_{\bm{i}}\sigma_0\right) c_{\bm{i}}
\end{gathered}\label{eq2_12}
\end{equation}
with $\tilde{v}^i_{\bm{i}}$ distributing uniformly in $[\kappa-W/2,\kappa+W/2]$ to the lattice model of the non-Hermitian Chern insulator $h_1(\bm{k})$ and calculate the transmission coefficient $T$ as a function of $W$ (in the unit of hopping energy $t$) for the same parameters as those for Fig.~3 in the main text. The results are shown in Fig.~\ref{fig_reply_2}. Since a global loss is effectively introduced through the disorders, chiral edge states become non-Hermitian such that $\langle T\rangle<1$.\par

\subsection*{S3: Topological Hermitian laser mode in a laser cavity network}

Experimentally~\cite{s_bandres_science_2018} and theoretically~\cite{s_gharari_science_2018}, it is shown that a laser cavity network possesses topologically protected chiral boundary modes in the presence of non-Hermicity. We consider a Haldane design of the laser network where the laser resonators are arranged in a honeycomb lattice. Each resonator is coupled to its nearest-neighbours with a real coupling constant $t_1$ and to its next-nearest-neighbours with a complex coupling constant $t_2\exp[i\phi]$, where $\phi$ is the Haldane flux parameter~\cite{s_gharari_science_2018}. The dynamics of the laser field reads
\begin{widetext}
\begin{equation}
\begin{gathered}
\left\{
\begin{array}{c}
i\dfrac{\partial a_{\bm{n}}}{\partial t}=\dfrac{\partial H}{\partial a^\ast_{\bm{n}}}=\omega_0 a_{\bm{n}}+t_1\sum_{\langle\bm{nm}\rangle}b_{\bm{m}}+t_2e^{i\phi}\sum_{\langle\langle \bm{nm} \rangle\rangle}a_{\bm{m}}-i\gamma a_{\bm{n}}+\dfrac{ig_a \mathbb{P}}{1+|\Psi|^2/I_{\text{sat}}}a_{\bm{n}}\\
\quad\\
i\dfrac{\partial b_{\bm{n}}}{\partial t}=\dfrac{\partial H}{\partial b^\ast_{\bm{n}}}=\omega_0 b_{\bm{n}}+t_1\sum_{\langle\bm{nm}\rangle}a_{\bm{m}}+t_2e^{-i\phi}\sum_{\langle\langle \bm{nm} \rangle\rangle}b_{\bm{m}}-i\gamma b_{\bm{n}}+\dfrac{ig_b \mathbb{P}}{1+|\Psi|^2/I_{\text{sat}}}b_{\bm{n}}.
\end{array}
\right..
\end{gathered}\label{eq2_1}
\end{equation}
\end{widetext}
Here, $a_{\bm{n}}$ and $b_{\bm{n}}$ are the laser field amplitudes at a site $\bm{n}=(x_{\bm{n}},y_{\bm{n}})$ of A and B sub-lattices, respectively. $\langle\bm{nm}\rangle$ and $\langle\langle\bm{nm}\rangle\rangle$ denote the nearest-neighbour sites and the next-nearest-neighbor sites, respectively. $\omega_0$ is the resonance frequency of the resonator. $\gamma$ represents the linear loss of a resonator. The last terms in Eq.~\eqref{eq2_1} represent optical gains via stimulated emission that is inherently saturated ($I_{\text{sat}}$). $\mathbb{P}$ is the pump profile, and $g_{a,b}$, the gain parameters of sub-lattice $A$ and $B$, depend on the pump intensity that can be tuned. Hereafter, we set 
\begin{equation}
\begin{gathered}
\tilde{g}_{a,b}=\dfrac{g_{a,b} \mathbb{P}}{1+|\Psi|^2/I_{\text{sat}}}.
\end{gathered}\label{eq2_2}
\end{equation}
\par

From Eq.~\eqref{eq2_1}, we derive the effective non-Hermitian Hamiltonian of the topological insulator laser:
\begin{equation}
\begin{gathered}
H=\sum_{\bm{n}}\left[\left(-i\gamma+i\tilde{g}_a\right)|a_{\bm{n}}|^2+\left(-i\gamma+i\tilde{g}_b\right)|b_{\bm{n}}|^2\right]+H_{\text{Haldane}}.
\end{gathered}\label{eq2_3}
\end{equation}
$H_{\text{Haldane}}$ is the standard Haldane Hamiltonian~\cite{s_haldane_prl_1988} that depends on the resonance frequency of a single resonator. In the absence of disorders, $H$ can be block diagonalized in the momentum space
\begin{equation}
\begin{gathered}
H=\sum_{\bm{k}}
\begin{bmatrix}
a^\ast_{\bm{k}} & b^\ast_{\bm{k}}
\end{bmatrix}h(\bm{k})
\begin{bmatrix}
a_{\bm{k}} \\
b_{\bm{k}}
\end{bmatrix},
\end{gathered}\label{eq2_4}
\end{equation}
where
\begin{equation}
\begin{gathered}
h(\bm{k})=h_0\sigma_0+\bm{h}\cdot\bm{\sigma}
\end{gathered}\label{eq2_5}
\end{equation}
with
\begin{equation}
\begin{gathered}
\left\{
\begin{array}{ccc}
h_0 & = & 2t_2\cos\phi\sum_{i=1,2,3} \cos[\bm{k}\cdot\bm{v}_i]+i\left[-\gamma+\dfrac{1}{2}(\tilde{g}_a+\tilde{g}_b)\right] \\
\quad\\
h_1 & = & t_1\sum_{i=1,2,3}\cos [\bm{k}\cdot\bm{u}_i] \\
\quad\\
h_2 & = & -t_1\sum_{i=1,2,3}\sin[\bm{k}\cdot\bm{u}_i] \\
\quad\\
h_3 & = & 2t_2\sin\phi\sum_{i=1,2,3}\sin[\bm{k}\cdot\bm{v}_i]+i\dfrac{\tilde{g}_a-\tilde{g}_b}{2}
\quad\\
\end{array}
\right..
\end{gathered}\label{eq2_6}
\end{equation}
Here, $\bm{u}_1=(\sqrt{3}/2,1/2),\bm{u}_2=(-\sqrt{3}/2,1/2),\bm{u}_3=-(\bm{u}_1+\bm{u}_2)$, and $\bm{v}_1=\bm{u}_2-\bm{u}_3,\bm{v}_2=\bm{u}_3-\bm{u}_1,\bm{v}_3=\bm{u}_1-\bm{u}_2$. The Hermitian part of Eq.~\eqref{eq2_5} supports chiral edge states for $t_2\neq 0$ and $-\pi<\phi<\pi$. If the global gain/loss is prohibited, say
\begin{equation}
\begin{gathered}
\left[-\gamma+\dfrac{1}{2}(\tilde{g}_a+\tilde{g}_b)\right]=0,
\end{gathered}\label{eq2_7}
\end{equation}
the chiral edge states are Hermitian. 
\par

\begin{widetext}

\begin{figure}[htbp]
\centering
\includegraphics[width=1\textwidth]{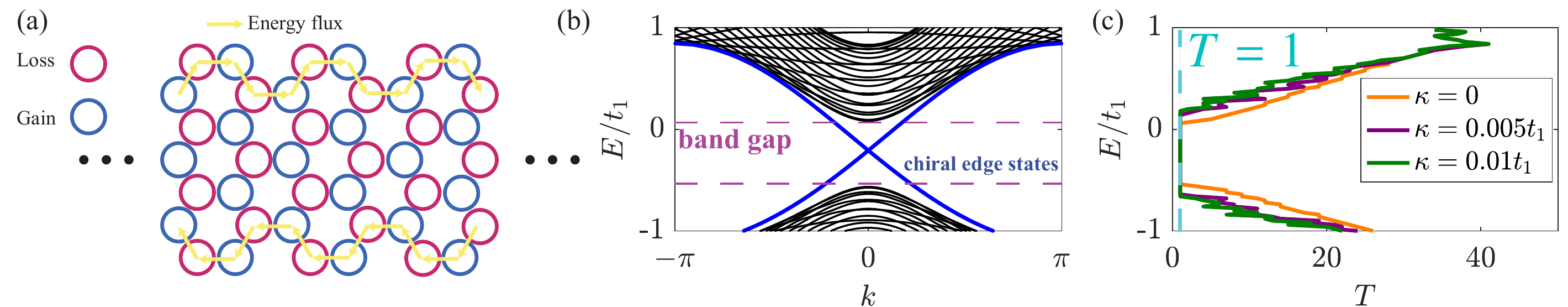}
\caption{(a) Geometry of a topological laser based on the Haldane design: A honeycomb lattice of coupled microring resonators with armchair edge in the $y$-direction. The output couplers (not shown here) are attached at the $x$-direction. A topological laser mode with unidirectional energy flux (yellow arrows) appears at boundaries. (b) Band spectrum of a passive (neither gain nor loss) topological insulator laser of the geometry shown in (a) [we have applied periodic boundary condition in the $x-$direction]. Here, $t_2=0.1t_1,\phi=\pi/5,\kappa=0$. (c) Quasi-particle energy $E$ (real part) v.s. transmission coefficient $T$ of laser modes in the presence of gain and loss given by Eq.~\eqref{eq2_8} for $\kappa=0.005t_1,0.01t_1$. As a comparison, those of the passive topological insulator laser are also plotted. Cyan dash line locates $T=1$.}
\label{fig_reply_1}
\end{figure}

\end{widetext}

To confirm above results, we choose 
\begin{equation}
\begin{gathered}
\tilde{g}_a=\gamma+\kappa,\tilde{g}_b=\gamma-\kappa
\end{gathered}\label{eq2_8}
\end{equation}
with $\kappa$ being a real positive number such that Eq.~\eqref{eq2_7} is satisfied. We specifically design the honeycomb lattice to be an armchair ribbon in the $y$-direction as shown in Fig.~\ref{fig_reply_1}(a). In the topological phase, e.g., $t_2=0.1t_1,\phi=\pi/5$, spectrum of this armchair strip has a band gap with topologically protected chiral edge states winding around the boundary, see the band spectrum of a passive case (no gain and loss) shown in Fig.~\ref{fig_reply_1}(b). Once the non-Hermicity is turned on by increasing $\kappa$, the topological band gap does not close, and no topological phase transition occurs. During this process, the bulk states become non-Hermitian, but, importantly, as shown in Fig.~\ref{fig_reply_1}(c), the chiral edge states remain Hermitian that is characterized by quantized transmission coefficient $T=1$. This means that the edge lasing modes in our design have uniform intensity along the propagating path.
\par
\quad\par

\end{document}